\documentstyle[preprint,aps]{revtex}
\begin{document}
\draft
\preprint{}
\title{ Doublet-Singlet Oscillations \\
          and\\
Dark Matter Neutrinos}
\author{ Nobuchika Okada 
 \thanks{e-mail: n-okada@phys.metro-u.ac.jp}
\thanks{JSPS Research Fellow}}
\address{Department of Physics, Tokyo Metropolitan University,\\
         Hachioji-shi, Tokyo 192-03, Japan}
\date{\today}
\preprint{
\parbox{4cm}{
\baselineskip=12pt
TMUP-HEL-9606\\
April, 1996\\
\hspace*{1cm}
}}
\maketitle
\vskip 2.5cm
\begin{center}
{\large Abstract}
\vskip 0.7cm
\begin{minipage}[t]{14cm}
\baselineskip=19pt
\hskip4mm
We examine the `singlet majoron model' first introduced 
by Chikashige, Mohapatra and Peccei 
as a simple extension of the standard model with massive Majorana neutrinos.
We can explain both the solar and the atmospheric neutrino deficits 
by the oscillations between electroweak doublet and singlet neutrinos 
without flavor mixing. 
Furthermore, while some light neutrinos can be the hot dark matter, 
tau neutrino with mass of $8.9$$-$$24 \rm{MeV}$ 
can be the cold dark matter through the interaction 
with the majoron.   
Thus, we can simultaneously explain the solar neutrino deficit, 
the atmospheric neutrino anomaly, and the cold and hot dark matters 
only with the Majorana neutrinos. 
\end{minipage}
\end{center}
\newpage
\def\barr{\begin{eqnarray}}
\def\earr{\end{eqnarray}}
There are several observations 
which can be explained if the neutrinos are massive. 
The solar electron neutrino deficit \cite{solar} 
can be understood 
by the neutrino oscillation phenomena due to the non-zero mass 
difference and the flavor mixing between neutrinos. 
There exist two types of solutions to the solar neutrino deficit:  
one is the oscillation with the Mikheyev-Smirnov-Wolfenstein (MSW) 
mechanism \cite{msw} 
inside the sun, the other is the solution of the vacuum oscillation   
from the sun to the earth \cite{vacuum}. 
The atmospheric neutrino anomaly \cite{atm}  
is the observations of the deficit of muon neutrino 
relative to the electron neutrino both of which are produced in the atmosphere. 
This also can be explained by the neutrino oscillation phenomena.  
If the sum of the mass of all neutrinos is $5$$-$$7$ eV, 
the neutrinos can play the role of the hot dark matter  
in the cold plus hot dark matter models \cite{chdm},  
which have good agreement with the observations of 
the matter distribution in the universe.   

The two observations of the neutrino deficits have been examined 
by the analysis of the neutrino oscillation 
with two flavor mixing or three flavor mixing scheme \cite{hlfl}. 
These analysis requires two mass squared differences:  
one is $\Delta m_{\odot}^2  \simeq 10^{-5} {\rm{eV}}^2$ 
($\Delta m_{\odot}^2  \simeq 10^{-10}{\rm{eV}}^2$) 
for the solar neutrino deficit with (without) the MSW mechanism,  
the other is $\Delta m_{\oplus}^2  \simeq 10^{-2}{\rm{eV}}^2$ 
for the atmospheric neutrino anomaly. 
Considering all of the observations mentioned above, 
it is required for three neutrinos of different flavors 
to have nearly degenerate masses of 
a few eV. 
However, if the neutrinos are the Majorana particles, 
this mass spectrum is excluded by the experiments of 
the neutrino-less double beta decay \cite{dbeta}, by which 
the effective electron neutrino mass is constrained as 
$\langle m_{\nu_e}\rangle < 1 \rm{eV}$. 

In this letter, 
we examine a model with the Majorana neutrino, 
called the `singlet majoron model',   
first considered by Chikashige, Mohapatra and Peccei \cite{cmp}.  
We show that the solar and the atmospheric neutrino deficits can be explained 
with oscillations between electroweak doublet and singlet neutrinos, but 
without flavor mixing. 
The existence of neutrinos as both the hot and cold dark matters 
is also shown. 

We extend the standard model by introducing three right-handed neutrinos 
and one electroweak singlet scalar.  
Since we assume the absence of the flavor mixing in the following, 
we can treat each generation separately. 
The Yukawa interactions for one generation are described by 
\barr
{\cal L}_{\rm{Yukawa}}=-g_{{}_Y} \overline{\nu _{L }}\Phi \nu _{R } 
-g_{{}_M } \overline{\nu_{R}{}^c}\phi \nu_{R} +h.c. \; \; ,\label{yukawa}
\earr
where $\Phi$ is the electric-charge neutral component of 
the Higgs field in the standard model, and 
$\phi$ is the electroweak singlet field.
The Dirac and the Majorana mass terms appear by 
the non-zero vacuum expectation values of these scalar fields.  
The mass matrix is given by 
\barr
\left[ \begin{array}{cc}
\hspace{0.4cm} 0 \hspace{0.3cm}  & \hspace{0.3cm} m_{{}_D} \hspace{0.4cm} \\   
\hspace{0.4cm} m_{{}_D} \hspace{0.3cm}  & \hspace{0.3cm} M \hspace{0.4cm}
\end{array}\right]
\label{mat}\; ,
\earr
where $m_{{}_D}=g_{{}_Y}\langle \Phi \rangle$ is the Dirac mass term, 
and $M=g_{{}_M}\langle \phi \rangle=g_{{}_M}v/\sqrt{2}$ 
is the Majorana mass term. 
Since the symmetry of the lepton number is spontaneously broken by
$\langle \phi \rangle \neq 0$,
a massless Nambu-Goldstone boson called majoron exists. 
For two mass eigenstates, the light one $\nu_{\ell}$ and  the heavy one $\nu_h$, 
we obtain the couplings of the neutrinos with the majoron from eq.(\ref{yukawa}): 
\barr
{\cal L}_{\chi \nu}= - \frac{i}{\sqrt{2}}\; \chi  \; \left[ 
\sin^2\theta \; \overline{\nu_{\ell}}\; i \gamma_{5}\; 
\nu_{\ell} 
-\sin\theta \cos \theta 
\left\{ \overline{\nu_{\ell}} \;  i\gamma_{5}\; \nu_h 
+ h.c.\right\}  
+\cos^2 \theta \; \overline{\nu_h} \; i \gamma_{5}\; \nu_h
\right]  \label{fmajo} \; \; ,
\earr
where the field $\chi$ is the majoron field defined 
by $\chi/\sqrt{2}=\rm{Im}\phi$, and $\theta$ is a mixing angle introduced by 
diagonalization of the mass matrix in eq.(\ref{mat}).  

Note that the oscillation between the electroweak doublet  
and singlet neutrinos is possible \cite{dsoscil}, 
since the mass matrix is not diagonal.  
In the following discussion, this type of oscillation is called 
the `doublet-singlet oscillation'. 
The information for the mass squared difference and the mixing angle 
is related to the values of the matrix elements in eq.(\ref{mat}). 
The small mixing angle ($\sin \theta \ll 1$) requires $m_{{}_D} \ll M$, 
called the see-saw type mass matrix \cite{seesaw}, and 
$M$ is fixed by the value of the mass squared difference, 
$M\simeq \sqrt{\Delta m^2}$. 
The almost pseudo-Dirac type mass matrix \cite{pDirac}, $m_{{}_D}\gg M$, 
is required by the large mixing angle ($\sin \theta \sim 1$), 
and the relation, $\Delta m^2\simeq 2 m_{{}_D} M$, is obtained. 

It is clear that the solar electron neutrino deficit and the atmospheric 
muon neutrino deficit can be explained by the `doublet-singlet oscillation', 
since experiments observe only the deficits, but not appearance of 
the converted partner through the oscillation. 
The solar neutrino deficit can be interpreted by  
the `doublet-singlet oscillation' in the first generation, and 
the atmospheric one can be interpreted by the same in the second generation.   
This type of the model of the neutrino oscillation is a kind of the model 
including the oscillation between the electroweak doublet neutrino 
and a `sterile' neutrino \cite{cm}. 
In our model, the physical meaning of the `sterile' neutrino is clear: 
it is the right-handed neutrino,  
which is introduced to generate the Majorana mass.  
Since we have little information for tau neutrino  
except for its existence and the upper bound on the mass 
$m_{\nu_{\tau}}< 24\rm{MeV}$ \cite{upper}, 
it is not needed to consider the oscillation in the third generation.  

However, since we have six neutrinos,  
we should consider the constraint on the number of neutrino 
species from the big bang nucleosynthesis (BBN) \cite{bbn}: 
$N_{\nu}\simeq 3$, where $N_{\nu}$ is the number of neutrino species 
which are in thermal equilibrium at the BBN era 
(temperature of the universe $\simeq 1\rm{MeV}$).   
It is known that, in the first generation, 
only the electroweak doublet neutrino contributes at the BBN era 
(see ref.[15] for brief discussion), 
if we take the small-angle MSW solution 
($\Delta m_{\odot}^2\simeq 10^{-5}$ and 
$\sin^22\theta_{\odot}\simeq 10^{-2}$) or the vacuum oscillation solution 
($\Delta m_{\odot}^2\simeq 10^{-10}$ and
$\sin^22\theta_{\odot}\simeq 1$) to 
the solar neutrino deficit 
\footnote
{
Considering the matter effect on the earth, 
the large-angle MSW solution ($\Delta m_{\odot} \simeq 10^{-5}$ 
and $\sin^22\theta_{\odot}\simeq 0.6$) 
with $\nu_e \rightarrow \nu_s$ (the `sterile' neutrino) oscillation 
is disfavored without cosmological discussion. 
This fact is pointed out by Hata and Langacker in ref.[6]. 
}. 
Then we take these solutions in the first generation. 
However, this is not the case in the second generation, 
since $\Delta m_{\oplus}^2\simeq 10^{-2} \rm{eV}$ and 
the large mixing angle, ${\sin }^22\theta_{\oplus} > 0.6 $,   
are required to explain the atmospheric neutrino anomaly 
(also see ref.[15]).   
Both two neutrinos in the second generation contribute to $N_{\nu}$. 
Then, there already exist three species of neutrinos 
which is in thermal equilibrium at the BBN era: 
the doublet neutrino in the first generation, and two neutrinos 
in the second generation.   
Thus, the energy density of the remaining two neutrinos 
in the third generation should be small at the BBN era. 

This situation can be realized in two ways.  
One is that neutrinos decay rapidly, and disappear until the time 
of the BBN era ($\simeq 1\rm{s}$).  
This can be applied to the heavy neutrino in the third generation, 
since it decays into the light neutrino and the majoron through the 
interaction in eq.(\ref{fmajo}).  
The other way is that neutrinos 
decouple from other particles in non-relativistic regime.    
This way should be applied to the light neutrino in the third generation, 
since it is stable.    

First, we discuss the case of the light neutrino.   
If it decouples in non-relativistic regime, 
its energy density is suppressed by the Boltzmann factor 
$e^{-m/T}\; (m >T)$, and becomes negligible, 
where $m$ and $T$ are the mass of the neutrino and 
the decoupling temperature, respectively.  
Since this suppression should works at the BBN era (1 MeV), 
$m>1 \rm{MeV}$ is required. 
However, this region of the mass of tau neutrino is cosmologically excluded 
\cite{taucosmo}, 
since the density parameter $\Omega$ in the present universe 
becomes too large, $\Omega \gg 1$. 
This is true, if we consider only the electroweak interaction.  
However, note that there is the interaction 
between neutrinos and the majoron.  
Carlson and Hall, and Kitazawa et al. \cite{chwe} pointed out that    
neutrinos can be the cold dark matter through the interaction.  
We investigate that the light neutrino in the third generation 
can really be the cold dark matter in the following.  

Let us consider the interaction between neutrinos in the third 
generation and the majoron. 
We assume that the mass matrix in the third generation 
is the see-saw type: $m_{{}_D}\ll M$ in eq.(2). 
Then, the light mass eigenstate 
($\nu_{\ell}$) and the heavy one ($\nu_h$) have masses 
$m_{\ell}\simeq m_{{}_D}^2/M$ and $m_h\simeq M$, respectively.  
The light neutrino is almost electroweak doublet state, or tau neutrino, and 
the heavy one is almost electroweak singlet state by the see-saw mechanism.  
Using $m_{\ell}$ and $m_h$, 
the couplings of the neutrinos with the majoron in eq.(\ref{fmajo}) 
are rewritten by 
\barr 
{\cal L}_{\chi \nu}\simeq - \chi  \; \left[
\left(\frac{m_{\ell}}{v}\right)\overline{\nu_{\ell}}\; i \gamma_{5}\;
\nu_{\ell} 
-\sqrt{\frac{m_{\ell}m_h}{v^2}}
\left\{ \overline{\nu_{\ell}} \;  i\gamma_{5}\; \nu_h 
+ h.c.\right\}  
+\left(\frac{m_h}{v}\right)\overline{\nu_h} \; i \gamma_{5}\; \nu_h
\right]  \label{majoron} \; \; ,
\earr
where the relation $m_h\simeq g_{{}_M}v/\sqrt{2}$ is used. 
 
The energy density of the cold dark matter 
in the present universe is given by 
\barr
\rho_{{}_{CDM}}\simeq 2.0\times 10^{-6}\;{\rm{GeV}}/{\rm{cm}}^3
\earr
in the cold plus hot dark matter models \cite{chdm}. 
These cosmological models agree very well  
with the observations of the matter distribution in the universe 
with the total density parameter $\Omega = 1$ and 
the Hubble constant 
$h\equiv H_0/100 ~\rm{km} ~\rm{s}^{-1} ~\rm{Mpc}^{-1}=0.5$. 

The relation between the mass $m_{\ell}(=m_{\nu_{\tau}})$ and $v$ 
is obtained by using the value of $\rho_{{}_{CDM}}$. 
The decoupling temperature $T_D$ is defined by \cite{kt}
\barr
n(T_D)\; \langle \sigma |v| {\rangle}_{T_D}  = H(T_D) \; , \label{dt}
\earr
where $n(T_D)$ is the number density of the tau neutrino at 
the decoupling temperature, 
$\langle \sigma |v|{\rangle}_{T_D} $ is the average value of the 
annihilation cross section of tau neutrino times relative velocity, 
and $H$ is the Hubble parameter.  
For the non-relativistic tau neutrino, $n(T_D)$ is approximately 
given by 
\barr
n(T_D) \simeq \frac{1}{\sqrt{2\pi^3}}\; x^{-\frac{3}{2}}\; 
e^{\frac{1}{x}}\; T_D^3
\; \; , 
\earr
where $x=T_D/m_{\nu_{\tau}}$. 
Considering the non-relativistic annihilation process of the tau neutrino, 
$\nu_{\tau} \nu_{\tau} \rightarrow \chi\chi$, we obtain 
\barr
\langle \sigma |v|{\rangle}_{T_D} \simeq \frac{1}{32 \pi}\;  
\frac{m_{\nu_{\tau}}T_D}{v^4}  \label{cross}
\earr
from eq.(\ref{majoron}). 
The  Hubble parameter $H$ is given by
\barr
H(T_D) = \left(\frac{8\pi^3g_*}{90}\right)^{\frac{1}{2}} \frac{T_D^2}
{M_P}\; \; ,
\earr
where $ M_P\simeq 1.2 \times 10^{19}\rm{GeV}$ is the Planck mass, 
and $g_*$ is the  total degrees of freedom 
of all particles in thermal equilibrium (we set $g_{*} =43/4+1$). 
The energy density of the tau neutrino in the present universe 
is given by 
\barr 
\rho_{\nu_{\tau}}=m_{\nu_{\tau}}\; n(T_D)\; 
\left(\frac{T_0}{T_D}\right)^3 \; \; , 
\label{energy} 
\earr
where $T_0\simeq 1.9 \rm{K}$ is the temperature of the tau neutrino 
at present. 
From eqs.(\ref{dt})-(\ref{energy}), and the condition 
$\rho_{\nu_{\tau}}= \rho_{{}_{CDM}}$, 
we can obtain the relation between $m_{\nu_{\tau}}$ and $v$. 
This relation is shown in Table I together with 
the contribution of the tau neutrino at the BBN era as another species, 
$\Delta N_{\nu} (=N_{\nu}-3)$. 
Considering  the experimental upper bound on the mass of tau neutrino,  
$m_{\nu_{\tau}}< 24\rm{MeV}$, 
and the BBN constraint (we take $\Delta N_{\nu}\leq 0.01$), 
the region, $ m_{\nu_{\tau}}\simeq 8.9$$-$$24 \rm{MeV}$ and 
$v\simeq2.7$$-$$4.3 {\rm{GeV}}$ is allowed as the cold dark matter. 

The heavy neutrino $\nu_h$ rapidly decay  
into the light neutrino (the tau neutrino) and the majoron. 
From eq.(\ref{majoron}), the life time of $\nu_h$ is described by 
\barr
\tau \simeq \frac{32\pi }{g_{{}_M}^2 m_{\nu_{\tau}}} \label{life}\; \; .
\earr
Substituting our result $m_{\nu_{\tau}}=8.9$$-$$24 \rm{MeV}$ into 
above equation, 
$\tau < 7.4\times 10^{-21}/g_{{}_M}^2\; {\rm{s}}$. 
The life time is far shorter than the age of the universe at the BBN era 
($ \simeq 1 \rm{s} $), unless $ g_{{}_M}$ is extremely small. 
Therefore, the heavy neutrino disappears until the time at the BBN era.  

The neutrinos in the first and second generation 
can be the hot dark matter.  
Since the mass squared differences required the solar and the atmospheric
neutrino deficits are far smaller than the mass scale 
of the hot dark matter, neutrinos as the hot dark matter 
have nearly degenerate masses.  
Two cases can be considered.  
One is that the two neutrinos in the second generation are
the hot dark matter with mass $\simeq 3 \rm{eV}$,  
if the small-angle MSW solution is taken in the first generation. 
In this case, the two neutrinos in the first generation have masses, 
$\sin^2 \theta \sqrt{\Delta m_{\odot}^2}$ and $\sqrt{\Delta m_{\odot}^2}$, 
respectively. 
The other case is that the neutrinos in both the first and second generations
are the hot dark matter with mass $\simeq 2 \rm{eV}$, 
if the solution of the vacuum oscillation is taken. 
These mass spectra in two generations are shown in Table II. 

There is no conflict with the experiments of 
the neutrino-less double beta decay, 
even if we take the vacuum oscillation solution 
and the mass $\simeq 2 \rm{eV}$ 
in the first generation. 
Note that the mass matrix in eq.(2) 
is almost pseudo-Dirac type, $m_{{}_D} \gg M$.  
Since we ignore the CP violating phase, 
the two mass eigenstates in the first generation 
have opposite CP eigenvalues:  
$\eta_{\pm}=\pm 1$. 
Considering that the solution of the vacuum oscillation requires 
the large mixing angle ($\theta_{\odot}\simeq \pi/4$), 
the effective electron neutrino mass is described by
\barr
\langle m_{\nu_e} \rangle \simeq \frac{1}{2}
|m_+ + \eta_+ \eta_- m_-|   \; \; , 
\earr
where $\eta_{\pm}=\pm 1$ are the CP eigenvalues, and 
$m_{\pm}\simeq m_{{}_D} \pm M/2$ are the mass eigenvalues. 
Then, we obtain 
\barr
\langle m_{\nu_e} \rangle \simeq \frac{M}{2} \simeq 
\frac{\Delta m_{\odot}^2}{4m_{{}_D}}
\simeq \frac{10^{-10}({\rm{eV}}^2)}{4\times 3({\rm{eV}})} 
\ll 1{\rm{eV}} \; \; ,
\earr
where the relation $M\simeq \Delta m_{\odot}^2 /2m_{{}_D}$ 
is used. 

Here, we must discuss the phenomena caused by the existence of 
the majoron. 
Although we showed that 
the number of neutrinos which exist at the BBN era is three, 
the majoron is in the thermal equilibrium at the BBN era and 
contribute $\Delta N_{\nu}=0.57$ as the additional species. 
Thus, our model results $N_{\nu}=3.57$.  
There are diverge BBN constraints obtained by many authors \cite{bbn}: 
$N_{\nu}<2.6$$-$$3.9$. 
Our result $N_{\nu}=3.57$ lies in this region, 
and is cosmologically allowed. 

The astrophysical bounds on the `singlet majoron model' should also  
be considered.  
The most restrictive constraint is obtained by the observations 
of neutrinos from the supernova 1987A \cite{sn1987}. 
These observations conclude that the gravitational binding energy is 
released almost by the emission of neutrinos. 
Thus, the energy release by other exotic particles 
is constrained smaller than 
that by neutrinos.  
Considering the majoron emission from the supernova, 
the constraint on the parameters in our model is obtained in two cases.  
One is the case in which the electroweak singlet Higgs boson,   
defined by $\sqrt{2}{\rm{Re}}\phi$, have mass less than the temperature of 
the core of the supernova ($T_{\rm{core}}=30$$-$$70{\rm{MeV}}$).  
The forbidden region is given by \cite{sn1} 
\barr
2\times 10^{-8} <\frac{m_{\nu_{\tau}}({\rm{MeV}})}
{v({\rm{GeV}})} < 3\times 10^{-7} \; \; . \label{ex1}
\earr
On the other hand, 
if the mass of the singlet Higgs boson 
is larger than the temperature of the core, 
the forbidden range is given by \cite{sn2} 
\barr 
2.3 \times 10^{-5} < \left( 
\frac{m_{\nu_{\tau}}}{{\rm{MeV}}}\right) \left(
\frac{{\rm{GeV}}}{v} \right)^2 
< 3.3\times 10^{-3} \; \; . \label{ex2} 
\earr
The region shown in Table I is outside these forbidden regions. 
Therefore, our model is astrophysically allowed. 

Next we consider the effect due to the interactions of the singlet neutrinos 
in the first and second generations with the majoron. 
Since the values of $M$ in the first and second generations are given by 
$M\simeq \sqrt{\Delta m_{\odot}^2}$ and $\sqrt{\Delta m_{\oplus}^2}$, respectively,  
the coupling constants of these neutrinos with the majoron is 
extremely small: 
$g_{{}_M}\simeq \sqrt{\Delta m_{\odot}^2}/v$ in the first generation, and  
$g_{{}_M}\simeq \sqrt{\Delta m_{\oplus}^2}/v $ 
in the second generation. 
Such an extremely weak interaction cannot affect in any 
cosmological or astrophysical observation.  

We would like to comment on the LSND experiment \cite{lsnd}. 
This experiment may be the first direct observation of the 
neutrino oscillation with 
$\overline{\nu_{\mu}} \rightarrow \overline{\nu_{e}}$.   
Since the LSND experiment is the type of the `appearance' experiment, 
it is clear that our model cannot explain this experiment. 
However, the explanation of the LSND result can be included, 
if we extend our model, 
and introduce flavor mixing between the first and the second generations. 
This extended model is the same, in form, as the models of ref.\cite{cm},  
in which the `sterile' neutrino is introduced, 
and the mixings among three flavor neutrinos 
and the `sterile' neutrino are investigated.  
However, note that our model is very restrictive, 
since there are only three mixing angles: 
one angle related to flavor mixing, and two angles corresponding to 
the mixing between the doublet and singlet neutrinos in the 
first and the second generations.  

Finally, we would like to mention the future solar neutrino experiments. 
The presence of the `doublet-singlet oscillation' in the first generation 
will be revealed 
in the future SNO \cite{sno} and Super-Kamiokande \cite{s-k} experiments 
as is pointed out by Bilenky and Giunti \cite{bg}. 
If the electroweak doublet neutrino converts to singlet one, 
the deficit of total flux of the solar neutrino is observed. 
The discovery of the `doublet-singlet oscillation' is  
a direct evidence of new physics beyond the standard model. 

In conclusion, 
we examine the `singlet majoron model' 
first introduced by Chikashige, Mohapatra and Peccei 
as a simple extension of the standard model with massive Majorana neutrinos.
In this model, 
we can explain the solar and the atmospheric neutrino deficits 
by the `doublet-singlet oscillations' without flavor mixing.  
Furthermore, while some light neutrinos can be the hot dark matter, 
tau neutrino with mass of $8.9$$-$$24 \rm{MeV}$ 
can be the cold dark matter through the interaction 
with the majoron.   
Thus, we can simultaneously explain 
the solar neutrino deficit, the atmospheric 
neutrino anomaly, and the cold and hot dark matters 
only with the Majorana neutrinos. 
The presence of the `doublet-singlet oscillation' 
(in the first generation) will be revealed 
in future SNO and Super-Kamiokande solar neutrino experiments. 

The author would like to thank Noriaki Kitazawa for numerous discussions, 
comments, and encouragements.  
This work was supported in part by the Grant in Aid for Scientific Research from 
the Ministry of Education, 
Science and Culture and the Research Fellowship of the Japan Society for the 
Promotion of Science for Young Scientists. 
%

%
\begin{table}
\caption{The relations among $v$, $m_{\nu_{\tau}}$, and $\Delta N_{\nu} $}
\begin{center}
\begin{tabular}{ccc}
\makebox[50mm]{$v \; (\rm{GeV})$}  & 
\makebox[50mm]{$m_{\nu_{\tau}}\; (\rm{MeV})$}  & 
\makebox[50mm]{$\Delta N_{\nu} $}     \\ \hline 
7  & 67 & $1.6\times 10^{-5}$  \\
6   & 48  & $1.6\times 10^{-5}$  \\
5   &  33 & $1.6\times 10^{-5}$  \\   
4  &   20   & $1.7\times 10^{-5}$   \\  
3  &  11  &   $2.2\times 10^{-3}$ \\
2  &  4.6 &  0.21  \\
1  &  1.0 &  0.92  \\ 
\end{tabular}
\end{center}
\end{table}
\begin{table}
\caption{The mass spectra of neutrinos in the first and second generations}
\begin{center}
\begin{tabular}{ccc}
\makebox[50mm]{solution}  &
\makebox[50mm]{first generation}  &
\makebox[50mm]{second generation}     \\ \hline
small-angle MSW & $m_{\ell} \simeq 8\times 10^{-6}$ eV 
& $m_{\ell} \simeq m_h \simeq 3 $ eV \\ 
  & $m_h \simeq  3\times 10^{-3}$ eV &   \\  \hline
vacuum oscillation  &  $m_{\ell} \simeq m_h \simeq 2$ eV 
& $m_{\ell} \simeq m_h \simeq  2$ eV  \\
\end{tabular}
\end{center}
\end{table}
\end{document}